\pdfoutput=1
\documentclass[11pt,a4paper]{article}
\usepackage{amsmath,amssymb}
\usepackage{braket}
\usepackage{url}
\usepackage{mathtools}
\usepackage{amsmath}              
\usepackage{graphics,graphicx,epsfig,ulem} 
\usepackage{subfigure,subfig}
\usepackage{feynmp}
\usepackage{slashed}
\usepackage{color}

\DeclareMathAlphabet{\mathbold}{U}{zeur}{b}{n}

\renewcommand\[{\left[}
\renewcommand\]{\right]}

\def\beq{\begin{equation}}
\def\eeq{\end{equation}}
\def\[{\begin{equation}}
\def\]{\end{equation}}

\begin{document}
\numberwithin{equation}{section}

\title{
{\normalsize  \mbox{}\hfill IPPP/17/30}\\
\vspace{2.5cm}
\Large{\textbf{Higgsplosion: 
Solving the Hierarchy Problem via rapid decays
of heavy states into multiple Higgs bosons}}}

\author{Valentin V. Khoze 
 and Michael Spannowsky\\[4ex]
  \small{\it Institute for Particle Physics Phenomenology, Department of Physics} \\
  \small{\it Durham University, Durham DH1 3LE, United Kingdom}\\
  [1ex]
  \small{\tt valya.khoze@durham.ac.uk \& michael.spannowsky@durham.ac.uk}\\
  [0.8ex]
}

\date{}
\maketitle

\begin{abstract}
 \noindent We introduce and discuss two inter-related mechanisms operative in the electroweak sector of the Standard Model at high energies. Higgsplosion, the first mechanism, occurs at some critical energy in the 25 to $10^3$ TeV range, and leads to an exponentially growing decay rate of highly energetic particles into multiple Higgs bosons. We argue that this a well-controlled non-perturbative phenomenon in the Higgs-sector which involves the final state Higgs multiplicities $n$ in the regime $n \lambda \gg 1$ where $\lambda$ is the Higgs self-coupling. If this mechanism is realised in nature, the cross-sections for producing ultra-high multiplicities of Higgs bosons are likely to become observable and even dominant in this energy range.
 At the same time, however, the apparent exponential growth of these cross-sections at even higher energies will be tamed and automatically cut-off by a related Higgspersion mechanism. As a result, and in contrast to previous studies, multi-Higgs production does not violate perturbative unitarity.
 Building on this approach, we then argue that the effects of Higgsplosion alter quantum corrections from very heavy states
 to the Higgs boson mass. Above a certain energy, which is much smaller than their masses, these states would rapidly decay into 
 multiple Higgs bosons. The heavy states become unrealised as they decay much faster than they are formed. 
 The loop integrals contributing to the Higgs mass will be cut off not by the masses of the heavy states, but by the 
 characteristic loop momenta where their decay widths become comparable to their masses. Hence, the 
 cut-off scale would be many orders of magnitude lower than the heavy mass scales themselves,  
 thus suppressing their quantum corrections to the Higgs boson mass.
\end{abstract}

\bigskip
\thispagestyle{empty}
\setcounter{page}{0}

\newpage


\section{Introduction}
\label{sec:intro}
\medskip

The recent discovery of a light Higgs boson at the Large Hadron Collider (LHC) \cite{Aad:2012tfa, Chatrchyan:2012xdj} constitutes an outstanding success of the Standard Model (SM) of particle physics. Before its discovery, the presence of a light scalar boson with a mass within the reach of the LHC was predicted, to ensure unitarity in scattering processes between longitudinal gauge bosons. While the SM is certainly an incomplete theory of nature, it fails to explain the observed matter-antimatter asymmetry and it does not provide a cold dark matter candidate, it is widely believed that the Higgs boson interactions with all other SM particles renders it a self-consistent theory, up to very high energy scales. As such it is currently arguably impossible to point to a specific energy scale at which the SM has to be augmented by new physics to explain fundamental questions in nature.

Yet, the Higgs boson, as a light elementary scalar particle, suffers from a so-called fine-tuning problem. Quantum corrections are involuntarily dragging the Higgs boson mass to the new physics mass scale $m_{\rm new}$, {\it viz} $m_h^2 \simeq m_0^2+\delta m_{\rm new}^2$. In order to obtain the observed physical mass of $m_h \simeq 125$ GeV the bare parameter of the theory $m_0$ has to be increasingly precisely tuned, depending on how widely the electroweak scale is separated from the new physics scale. The guiding principle that parameters of our quantum field theory should not have to be unnaturally precisely tuned is currently our strongest argument for the existence of a new physics scale, not too far away from the electroweak scale. Popular ways to avoid the Hierarchy problem altogether are supersymmetric and composite Higgs models, which each however have their own so-called little Hierarchy problems.

\medskip

Looking beyond $2 \to 2$ scattering processes, which are unitarized due to tree-level cancellation effects between gauge and Higgs boson interactions, the SM might still be an inconsistent theory at energy scales as low as $\mathcal{O}(100)~\mathrm{TeV}$, as perturbative unitarity might be violated in $2 \to nh$ multi-Higgs boson production processes. At sufficiently high energies it becomes kinematically possible to produce high multiplicity final states with $n\ggg 1$ particles in a weakly interacting theory. It was pointed out already more than a quarter of a century ago in Refs.~\cite{Cornwall:1990hh,Goldberg:1990qk} 
that the factorial growth in $n$ can
arise from the large numbers of Feynman diagrams contributing
to the scattering amplitude ${\cal M}_{n}$ at large $n$. 
This reasoning works in any quantum field theory where there is no destructive interference between Feynman diagrams in 
computations of on-shell quantities, and is indeed the case in the scalar field theory with $\lambda \phi^4$-type
interactions \cite{Brown}, where tree graphs all have the same sign, and the leading-order high-multiplicity amplitudes indeed
acquire the factorial behaviour, 
${\cal M}_{n} \sim \lambda^{n/2}\, n!$. This observation, assuming that the amplitudes do not decay rapidly in moving off the 
multi-particle thresholds,  leads to
the factorial growth of the decay rates, $\Gamma_n \sim \lambda^n\, n! \times f_n({E})$, of highly energetic
states and signals that perturbation theory becomes effectively strongly coupled for $n > 1/\lambda$ 
 \cite{Voloshin:1992rr,LRST,Son,VVK2,VVK3} and
can result in sharply growing with energy high-multiplicity observables. 
For example, it was shown recently in
Refs.~\cite{VVK3,Degrande:2016oan} that such high multiplicity production processes may be within reach of a
future hadron collider at 100 TeV. Already at 50 TeV the perturbative cross-sections for 140 Higgs bosons are at picobarn level.

\medskip
In this work, we will address both short-comings of the SM discussed: the Hierarchy problem and the apparent  breakdown of perturbative unitarity in high multiplicity processes simultaneously using the {\tt Higgsplosion} mechanism. We will show that the sharply growing cross-sections actually prevent the violation of perturbative unitarity in multi-Higgs processes and further naturally tame the effect of quantum corrections to the Higgs boson mass.
The key point here is that the decay width is the imaginary part of the 2-point correlator, with the 
LSZ-amputated external lines. In a physical process, for example when the highly virtual Higgs boson is produced as an intermediate state in the gluon fusion process before decaying into a high-multiplicity final state,
the amplitude is of the form,
\[
{\cal M}_{gg\to h^*} \,\times\, \frac{i }{p^2\,-\,M_h^2+i\, M_h\, \Gamma(p^2)} 
\,\times\, {\cal M}_{h^* \to n\times h} \,,
\label{eq:amp1}
\]
where $M_h$ is the Higgs mass and $\Gamma(s)$ is the energy-dependent total width of the Higgs at
the scale $s$. It is of course the same rate as the the result of computing the phase-space integral in the cross-section,
\[
 \int d\Pi_n |{\cal M}_{h^* \to n\times h}|^2 \,=\, 2 M_h \times \Gamma_n(s) \,\, , \qquad
\Gamma = \sum_{n} \Gamma_n\,.
\]
Hence the cross-section of the single-Higgs-exchange process \eqref{eq:amp1} at high energies is schematically
of the form
\[ \sigma_n \,\sim\, \frac{\sqrt{s} \, \Gamma_n(s)}{s^2+ M_h^2\, \Gamma^2(s)} \,,
\]
and at asymptotic energies, where $\Gamma_n\to \infty$, is in fact consistent with unitarity.

\bigskip

The occurrence of sharply growing decay rates of highly energetic (or massive) initial states into high-multiplicity states of 
relatively soft Higgs bosons (and in all likelihood other massive vector bosons) will be called the Higgsplosion effect. As it effectively amounts to an exploding multi-particle decay width $\Gamma_n(s)$ 
of supermassive heavy states $X$, Higgsplosion must affect their propagators,
\[
\Delta_X (p)\,=\,  \frac{i }{p^2\,-\,M_X^2-i\, {\rm Im}\, \Sigma_X(p^2)} \,=\,
\frac{i }{p^2\,-\,M_X^2 \,+i \,M_X\Gamma_X(p^2)} \,,
\label{eq:pfin}
\]
appearing in the loops contributing to the quantum corrections to the Higgs mass.
If, due to the Higgsplosion mechanism, the decay width $\Gamma_X$ of the heavy particle into $n$ Higgs bosons
exceeds the heavy mass $M_X$ at the scale $\sqrt{s_\star}$ which is much smaller than $M_X$, then it will be the scale $\sqrt{s_\star}$
rather than $M_X$ which will provide the cut-off of the loop integrals in the self-energy contributions to the Higgs mass.
Our central point is that purely on dimensional grounds, the Hierarchy problem for the Higgs mass is reduced by
a positive power of the factor of
$\frac{s_{\star}}{M_X^2} \ll 1$.

\bigskip

This article is organised as follows: In Sec.~\ref{sec:2} we review briefly how off-shell momenta enter the propagator and decay width of a scalar particle. We derive  the scaling behaviour for the dimensionless quantity $\mathcal{R}$ due to Higgsplosion in Sec.~\ref{sec:3}. In Sec.~\ref{sec:4} we introduce the Higgspersion mechanism, showing that perturbative unitarity is not violated in multi-Higgs production processes in the Standard Model. The connection between Higgsplosion and the dynamical taming of the Hierarchy problem we discuss in Sec.~\ref{sec:5}. In Sec.~\ref{sec:concl} we offer our conclusions.

\medskip
\section{Propagators and partial decay widths of massive fields}
\label{sec:2}
\medskip

We are interested in investigating quantum effects caused by steeply growing multi-particle decay rates 
of a highly virtual (or highly energetic) degree of freedom in the initial state above a certain critical energy. 
The decay widths enter the propagators of the relevant states, thus we start in this section with a brief review of the 
full propagator for a massive scalar.  In subsequent sections this will be used in our discussion of two
cases: the Higgs propagators 
appearing as intermediate states in high-energy cross-sections, and 
the ultra-heavy states contributing to the Higgs mass through loop effects.

\bigskip

Consider a simple quantum field theory of a single real scalar field $\phi$ described by the 
Lagrangian
\[ {\cal L}\,=\, \frac{1}{2} \partial^\mu \phi \,\partial_\mu \phi \,-\, \frac{1}{2} m_0^2 \, \phi^2 \,-\, {\cal L}_{\rm int} (\phi)\,,
\label{eq:toym}
 \]
where $m_0$ denotes the bare mass parameter and the interaction term ${\cal L}_{\rm int} (\phi)$ includes the usual 
renormalizable self-interactions of $\phi$, for example $ {\cal L}_{\rm int}=\,\frac{\lambda}{4!} \, \phi^4$ or 
${\cal L}_{\rm int}=\, \frac{\lambda}{4} (\phi^2-v^2)^2$. The Feynman propagator of $\phi$ is the Fourier transformation
of the 2-point Green function, and reads
\[ \Delta_\phi (p) \,=\, \int d^4x \, e^{i p\cdot x} \langle 0|T\left( \phi(x) \, \phi(0)\right)|0\rangle \,=\,
\frac{i}{p^2-m_0^2-\Sigma(p^2)+i\epsilon}\,,
\label{eq:dres}
\]
where $\Sigma(p^2)$ is the self-energy of $\phi$, i.e. $-i \Sigma(p^2)$ is the the sum of all one-particle-irreducable (1PI) 
diagrams contributing to the two-point function. It is related to the amplitude for a $1 \to 1$ particle scattering, 
${\cal M}(p\to p)$ via the LSZ reduction formalism, so that
\[ {\cal M}(p\to p) \,=\, - Z_\phi \, \Sigma(p^2)\,,
\label{eq:lsz}
\]
and $Z_\phi$ is the wave-function renormalization constant.
What we have on the
right hand side of Eq.~\eqref{eq:dres} is the resummed or dressed propagator since it can be Taylor expanded 
in terms of the bare propagators and the self-energy insertions,
\[
\frac{i}{p^2-m_0^2-\Sigma(p^2)}\,=\, \frac{i}{p^2-m_0^2} \,+\, 
 \frac{i}{p^2-m_0^2}\, \sum_{n=1}^\infty \left(-i \Sigma(p^2)\, \frac{i}{p^2-m_0^2}\right)^n
\,.
\label{eq:drexp}
\]
For simplicity, from now on, we are dropping the $i\epsilon$ factor in the propagators.

The physical (or pole mass) mass $m$ is then defined as the location of the pole in the full propagator of Eq.~\eqref{eq:dres}. It 
is the solution of the equation,\footnote{In our toy-model the particles are absolutely stable
near their mass-shell. The model contains only self-interactions of the field $\phi$ and the decays become kinematically allowed 
only at energies above the multi-particle mass-thresholds, i.e. $p^2> (2m)^2$. Hence the self-energy $\Sigma(p^2=m^2)$
contains no imaginary part below the higher-particles mass-thresholds, hence $\Sigma(p^2=m^2)= {\rm Re} \,\Sigma(p^2=m^2)$
and the pole in \eqref{eq:mdef} is on the real axis.}
\[
m^2 - m_0^2 -  \Sigma(m^2)\,=\, 0 \,\,, \quad {\rm or} \quad 
m^2\,=\, m_0^2 + {\rm Re}\, \Sigma(m^2)\,.
\label{eq:mdef}
\]
The meaning of the self-energy at the fixed scale $p^2=m^2$ is that it provides the shift to the bare mass,
${\rm Re}\, \Sigma(m^2)=\delta m^2,$ 
in order to obtain the observable and finite physical mass $m^2=m_0^2+\delta m^2.$

Using the equation \eqref{eq:mdef} for the physical mass we can represent the dressed propagator Eq.~\eqref{eq:dres} in the form,
\[\Delta_\phi (p)\,=\, \frac{i}{p^2-m^2-[\Sigma(p^2)-\Sigma(m^2)]}\,=\, \frac{i}{p^2-m^2}
\left(\frac{1}{1-\frac{d \Sigma}{dp^2}|_{p^2=m^2}+ {\cal O}(p^2-m^2)}\right)\,,\nonumber
\]
which in the limit $p^2\to m^2$ results in the well-known pole form of the propagator,
\[\left.\Delta_\phi (p)\right\vert_{p^2 \to m^2}\,\simeq\, \frac{i Z_\phi}{p^2-m^2}\,\,, \quad {\rm where} \quad
Z_\phi \,=\, \left(1 -\left.\frac{d \Sigma}{dp^2}\right\vert_{p^2=m^2}\right)^{-1}\,.
\label{eq:mass1}
\]
$Z_\phi$ is the field renormalization constant which already appeared in Eq.~\eqref{eq:lsz}.

\bigskip

In this paper we will be mostly interested in the kinematic regime(s) far away, i.e. far above or far below, from the single-particle mass shell region
$p^2 \simeq m^2$ of the propagator in Eq.~\eqref{eq:mass1}. In the case of the light stable field $\phi$ 
we are considering at present, the regime of interest is such that multi-particle decays with ultra-high multiplicities 
$n \gg 1/\lambda\gg 1$
can contribute to the propagator, and hence $p^2 \gtrsim (n m)^2 \gg m^2$. In this case the propagator is
described by the full expression of Eq.~\eqref{eq:dres}, and the self-energy contains a non-vanishing imaginary part.
Specifically we will concentrate on the scenarios where multi-particle decays of a virtual $\phi$ into
$n$-particle states, with ultra-high multiplicites $n$ lead to decay widths which grow sharply with energy 
$E=\sqrt{s}$ above some critical value $E_{\rm crit}$. If this scenario is realised in nature, one can enter
 the energy regime where ${\rm Im} \Sigma (s) \gg m^2$. This is the regime of interest we will concentrate on in this work.

In the single-field toy model of Eq.~\eqref{eq:toym} we are discussing at present, the particles described by the field $\phi$ 
are well-defined asymptotic states of mass $m$ and they are absolutely stable not too far above their single-particle mass threshold,
$m^2 \le p^2 < (2m)^2$. Indeed, we have assumed that $\phi$ interacts only with itself, and there are no interactions with lighter 
states in the Lagrangian.
This results in multi-particle thresholds at $p^2 \ge (nm)^2$ for $n=2,3,\ldots$ corresponding to $\phi \to n\times \phi$ decays
at energies $s \ge (nm)^2$ for $n\ge 2$. Thus, at around the single-particle mass-shell the decay width is zero, the propagator
is real-valued and contains only the pole term -- as indicated by Eq.~\eqref{eq:mass1}. However, at higher energy scales,
the multi-particle mass thresholds are reached resulting in the appearance of the imaginary part of $\Sigma(p^2)$ in the
full propagator on the right hand side of Eq.~\eqref{eq:dres}. 
For the full propagator we have
\begin{eqnarray}
\Delta_\phi (p) &=& \frac{i}{p^2-m^2-{\rm Re}[\Sigma(p^2)-\Sigma(m^2)]-i {\rm Im}\Sigma(p^2)} \nonumber\\
&=& \frac{i Z_\phi}{p^2-m^2-i Z_\phi\,  {\rm Im}\, \Sigma(p^2)} \,+\, \ldots
\label{eq:drexpImS}
\end{eqnarray}
In deriving this expression we Taylor-expanded the quantity
\[{\rm Re}[\Sigma(p^2)-\Sigma(m^2)] = {\rm Re} \left.\frac{d \Sigma}{dp^2}\right\vert_{m^2}(p^2-m^2) +
{\cal O}((p^2-m^2)^2)\,,\]
and used the definition of the wave-function renormalization constant \eqref{eq:mass1}. The dots on the right hand side
of Eq.~\eqref{eq:drexpImS} denote the contributions of higher order terms in the Taylor expansion of ${\rm Re}(\Sigma(p^2))$
which will aways assume to be subleading to the effects we want to study here and that they can be treated 
as higher-order corrections in pertrurbation theory.

We will thus use the following expression for the scalar field propagator 
\[ \Delta_\phi (p)\,\simeq\, 
 \frac{i Z_\phi}{p^2-m^2-i Z_\phi\,  {\rm Im}\, \Sigma(p^2)} \,=\,
\frac{i Z_\phi}{p^2-m^2 +i \,m \,\Gamma(p^2)} \,,
\]
where we traded the imaginary part of the self-energy for the energy-dependent decay width $\Gamma(p^2)$,
{\it cf.} Eq.~\eqref{eq:lsz},
\[ - Z_\phi\,  {\rm Im}\, \Sigma(p^2) \,=\, {\rm Im}\,  {\cal M}(p\to p) \,=\, m\,\Gamma(p^2) \label{eq:2.10}
\,,\]
with the decay width being determined by the partial widths of $n$-particle decays at energies
$s \ge (nm)^2$,
\[
\Gamma (s) \,=\, \sum_{n=2}^\infty \Gamma_n (s) \,\,, \qquad
\Gamma_n (s)\,=\,\frac{1}{2m} \int d\Pi_n |{\cal M}(1\to n)|^2 \,. \label{eq:2.11}
\]
${\cal M}$ is the amplitude for the $1^*\to n$ process and the integral is over the $n$-particle Lorentz-invariant
phase space.

\bigskip
In summary, for the UV-renormalised propagator $\Delta_{R} (p) \,= \, Z_\phi^{-1}$, 
we will use the following expression in terms of the pole mass $m^2$, the
renormalised self-energy $\Sigma_R(p^2)\,=\,Z_\phi\, \Sigma(p^2)$, or the physical width $\Gamma(p^2)$,
and the renormalised coupling constant(s),
\[ \Delta_R (p)\,=\, 
 \frac{i }{p^2-m^2-i\, {\rm Im}\, \Sigma_R(p^2)} \,=\,
\frac{i }{p^2-m^2 +i \,m\, \Gamma(p^2)} \,.
\label{eq:pfin2}
\]
All quantities in the expression above are UV-finite. The framework of using the propagator for the Higgs boson 
with the energy-dependent width as the correct description, applicable for all kinematic regions is widely used
in the literature, see e.g. Refs.~\cite{Seymour:1995np,Goria:2011wa},
and is consistent with our treatment.\footnote{In this paper we focus exclusively on multi-Higgs decays and
are not concerned with the decays of the Higgs 
into lighter SM particles below its mass threshold. These can be readily incorporated.}
In the following section we will concentrate on the decay width $\Gamma(s)$.

\medskip
\section{Multi-particle decay width of the Higgs boson}
\label{sec:3}
\medskip

We now consider the ultra-high multiplicity Higgsplosions of highly energetic
virtual particles in the Standard Model. Specifically, we will describe the main features of the mechanism
using a simplified model for the Standard Model Higgs boson in terms of a QFT of 
 a single real scalar field $h(x)$ with non-vanishing vacuum expectation value (VEV) $\langle h \rangle = v$,
\[{\cal L} \,= \, \frac{1}{2}\, \partial^\mu h \, \partial_\mu h\, -\,  \frac{\lambda}{4} \left( h^2 - v^2\right)^2
\,.
\label{eq:Lh}
\]
This theory is a reduction of the SM Higgs sector in the unitary gauge to a single scalar degree of freedom,
$h(x)$ which for our purposes we take to be stable, so there are no decays into fermions, and we have also decoupled 
all vector bosons etc. The physical VEV-less scalar $\varphi(x) = h(x)-v$,
describs the Higgs boson of mass $M_h = \sqrt{2\lambda}\,v$ and satisfies the classical equation arising from Eq.~\eqref{eq:Lh},
\[
-\, (\partial^\mu \partial_\mu +M_h^2)\, \varphi \,=\, 3\lambda v\, \varphi^2 \,+\,  \lambda\,\varphi^3.
\label{eq:varphi}
\]
The first step in our programme is to determine the multi-particle amplitudes describing the $1^* \to n$ transitions
of the highly virtual Higgs boson into $n$ non-relativistic Higgses at the leading order (i.e. tree-level) in perturbation theory. 
We take the 
bosons in the final state to be non-relativistic because we are interested in keeping the number of particles $n$ 
in the final state as large as possible, that is, near the maximum number allowed by the phase space, 
$n \lesssim n_{\rm max} = E/M_h$.
Such $n$-point amplitudes were studied in detail in scalar QFT in \cite{Brown,LRST} and were derived for the theory of Eq.~\eqref{eq:Lh} with spontaneous symmetry breaking in Ref.~\cite{VVK2},
\[
{\cal A}_{1^*\to n} (p_1 \ldots p_n)  \,=\, n!\,  (2v)^{1-n}\,\exp\left[-\frac{7}{6}\,n\, \varepsilon \right]\,,
\quad n\to \infty\,, \quad \varepsilon \to 0\,, \quad n\varepsilon = {\rm fixed}\,.
\label{eq:expsc}
\]
Note that the expression above is for the ${1^*\to n}$ current, and the conventionally-normalised 
amplitude ${\cal A}_{1^*\to n}$ is obtained from it by the LSZ amputation of the single off-shell incoming line,
\[
{\cal M}_{1\to n}  \,: =\, (s-M_h^2) \,\cdot\,{\cal A}_{1^*\to n} (p_1 \ldots p_n) \,\,.
\]
As indicated, these tree-level amplitudes are computed in the double-scaling limit with large multiplicities $n \gg 1$ 
and small non-relativistic energies of each individual particle, $\varepsilon \ll 1$, where 
\[
\varepsilon \,= \, \frac{\sqrt{s}-nM_h}{nM_h}\,\,\,=\,\,\,\, \frac{1}{n\, M_h}\, E_n^{\rm \, kin}\, \simeq\, 
\frac{1}{n}\,  \frac{1}{2 M_h^2} \, \sum_{i=1}^n \vec{p}_i^{\, \, 2} \,,
\label{eq:epsdef}
\]
so that the total kinetic energy per particle mass $n\varepsilon$ in the final state is fixed.
The first factor on the right-hand side of Eq.~\eqref{eq:expsc} corresponds to the tree-level amplitude 
(or more precisely a current with one incoming off-shell leg) computed on
the $n$-particle threshold, 
\[ {\cal A}^{\rm thr.}_{1\to n}= n!\,  (2v)^{1-n} = n!\,\left(\frac{\lambda}{M_h^2}\right)^\frac{n-1}{2}\,,
\]
or, equivalently, after the LSZ reduction of the incoming line,
\[ 
{\cal M}^{\rm thr.}_{1\to n}\,=\, n!\,(n^2-1) \, \frac{\lambda^\frac{n-1}{2}}{M_h^{n-3}}\,,
\]
which is an
exact expression for tree-level amplitudes valid for any value of $n$ \cite{Brown}. The kinematic dependence in Eq.~\eqref{eq:expsc} then produces in the non-relativistic limit an exponential form-factor which has an analytic dependence on the kinetic energy of the final state $n\varepsilon$.
But, importantly, the factorial growth $\sim \lambda^{n/2}\,  n!$ characteristic to the multi-particle amplitude on mass threshold remains.
Its occurrence can be traced back to the factorially growing number of Feynman diagrams at large $n$
\cite{Dyson:1952tj,Lipatov:1976ny,Brezin:1976vw} and the lack of destructive interference 
between the diagrams in the scalar theory.
We refer the reader to Refs.~\cite{Brown,LRST,VVK2} for more detail about these amplitudes.

The next step is to integrate the amplitudes in Eq.~\eqref{eq:expsc} over the $n$-particle phase-space at large $n$
(in the approximation where the outgoing particles are 
non-relativistic). The relevant dimensionless quantity describing the multi-particle processes is
\[ 
{\cal R}_n (s)\,:=\,\frac{1}{2M_h^2} \int d\Pi_n |{\cal M}(1\to n)|^2 \,, 
\label{eq:Rdef} 
\]
and the decay rates $\Gamma_n(s)$ and the cross-sections $\sigma_n(s)$ are obtained from 
${\cal R}_n (s)$ after an appropriate overall rescaling with $M_h$ and $s$.
Following in the steps of Refs.~\cite{Son,VVK2}, we obtain the characteristic  
exponential expression for the $1\to n$ particles rate ${\cal R}$ in the high-energy, high-multiplicity limit:
\begin{eqnarray}
\label{eq:Rtreesimp}
&&{\cal R}(\lambda; n,\varepsilon)\,=\, 
\exp \left[ n\left(\log \frac{\lambda n}{4} -1 \right) \,+\, \frac{3n}{2}\left(\log \frac{\varepsilon}{3\pi} +1 \right)
\, -\,\frac{25}{12}\,n\varepsilon \right] \,,
\\ \nonumber\\
&&\Gamma_n(s) \, \propto {\cal R}(\lambda; n,\varepsilon)\,\,, \qquad {\rm and} \qquad 
\sigma_n(s) \, \propto {\cal R}(\lambda; n,\varepsilon)\,. \nonumber
\end{eqnarray}
In particular, note that the ubiquitous factorial growth of the large-$n$ amplitudes  translates into the 
 $\frac{1}{n!} |{\cal M}_{n}|^2 \sim n! \lambda^n \sim e^{n\log(\lambda n)}$ factor in the rate ${\cal R}$ above.
 
To summarise our discussion so far, let us consider the multi-particle limit $n \gg 1$ and scale the center-of-mass energy 
$\sqrt{s}=E$ linearly with $n$, $E \propto n$, keeping the copupling constant small at the same time, $\lambda \ll 1$. 
It was pointed out first in Refs.~\cite{LRST,Son}, and then argued for extensively in the literature,
that in this limit the multi-particle rates have a characteristic exponential form,
\[
{\cal R} \,=\, e^{ \,n F(\lambda n, \,\varepsilon)} \,,
\quad {\rm for}\,\, n \to \infty\,,\,\, \lambda \to 0\,,\,\, \varepsilon ={\rm fixed}\,,
\label{eq:hg}
\]
where it is assumed that the high-multiplicity, weak-coupling limit above, the factor $\lambda n$ is held fixed, 
while the fixed value can be small or large (with the former case allowing for a perturbative treatment, while
the latter one requiring a large $\lambda n$ resummation of perturbation theory, somewhat reminiscent 
to the large $g^2 N_c$ 't Hooft coupling limit in gauge theories).
The quantity $\varepsilon$ is the average kinetic energy per particle per mass in the final state of Eq.~\eqref{eq:epsdef},
and $F(\lambda n, \varepsilon)$ is a certain a priori unknown function of two arguments.
At tree-level, the dependence on $\lambda n$ and $\varepsilon$,
factorises into individual functions of each argument,
\[
F^{\rm tree}(\lambda n, \,\varepsilon) \,=\, f_0(\lambda n)\,+\, f(\varepsilon)\,,
\label{eq:nFtree}
\]
and the two independent functions are given by the following expressions in the Higgs model of 
Eq.~\eqref{eq:Lh},  in complete agreement with the expression Eq.~\eqref{eq:Rtreesimp},
\begin{eqnarray}
\label{f0SSB}
f_0(\lambda n)&=&  \log\left(\frac{\lambda n}{4}\right) -1\,, 
\\
\label{feSSB}
f(\varepsilon)|_{\varepsilon\to 0}&\to& f(\varepsilon)_{\rm asympt}\,=\, 
\frac{3}{2}\left(\log\left(\frac{\varepsilon}{3\pi}\right) +1\right) -\frac{25}{12}\,\varepsilon\,.
\end{eqnarray}

One can further come up with various improvements in the understanding and control of the exponential behaviour
of the multi-particle rate. In particular, at tree-level the function $f_0(\lambda n)$ is fully determined, but the second function,
 $f(\varepsilon)$, characterising the energy-dependence of the final state, is determined by Eq.~\eqref{feSSB} only at small $\varepsilon$, i.e. near the multi-particle threshold. This point was addressed recently in Ref.~\cite{VVK3}
 where the function $f(\varepsilon)$ was computed numerically in the entire range  $0\le \varepsilon < \infty$. 
 
 What about the inclusion of loop corrections to the tree-level multi-particle rates above? 
 This has been achieved at the leading order in $\lambda n$ in Ref.~\cite{LRST}
 by resumming the one-loop correction to the amplitude on the multi-particle mass threshold 
 computed in Refs.~\cite{Voloshin:1992nu,Smith:1992rq}. The result is that the 1-loop correction 
 in the Higgs theory under consideration does not affect the factorial growth, but provides an
 exponential enhancement to the rate (though strictly speaking it is valid only at small values of $\lambda n$)
 and results in the modified expression for $f_0$, 
\[
f_0(\lambda n)^{\rm 1-loop} \,=\, \log\left(\frac{\lambda n}{4}\right) -1 \,+\, \sqrt{3}\, \frac {\lambda n}{4\pi}\,.
\label{eqnl}
\]

\bigskip

Of phenomenological interest is whether the multi-particle rates can become observable at certain energy scales and, at even
higher energies, exponentially large -- in the limit of near maximal kinematically allowed multiplicities. To answer this, it is required
 to resum the perturbation theory and address the large $\lambda n$ limit. Very recently, we have computed the exponential 
 rate in the $\lambda n \gg 1$ limit using the Landau WKB-based formalism, following the approach of Ref.~\cite{Son}. 
 These results will be reported in a forthcoming publication \cite{VVKinprog}. The correction 
to the tree-level rate in the non-relativistic regime is found to be of the form $\approx +3.02\, n\, \sqrt{\frac{\lambda n}{4\pi}}$. 
 
 As a result, the non-perturbatively corrected multi-particle rate in Eq.~\eqref{eq:Rtreesimp} becomes \cite{VVKinprog}
 \[ {\cal R} \,=\, 
\exp \left[ \frac{\lambda n}{\lambda}\, \left( 
\log \frac{\lambda n}{4} \,+\, 3.02\, \sqrt{\frac{\lambda n}{4\pi}}\,-\,1\,+\,\frac{3}{2}\left(\log \frac{\varepsilon}{3\pi} +1 \right)
\, -\,\frac{25}{12}\,\varepsilon 
\right)\right] \,. 
\label{eq:Rnp}
\]
 This expression is derived at small $\varepsilon$ and thus is supposed to hold in the non-relativistic limit.

 \begin{figure*}[t]
\begin{center}
\begin{tabular}{c}
\includegraphics[width=0.75\textwidth]{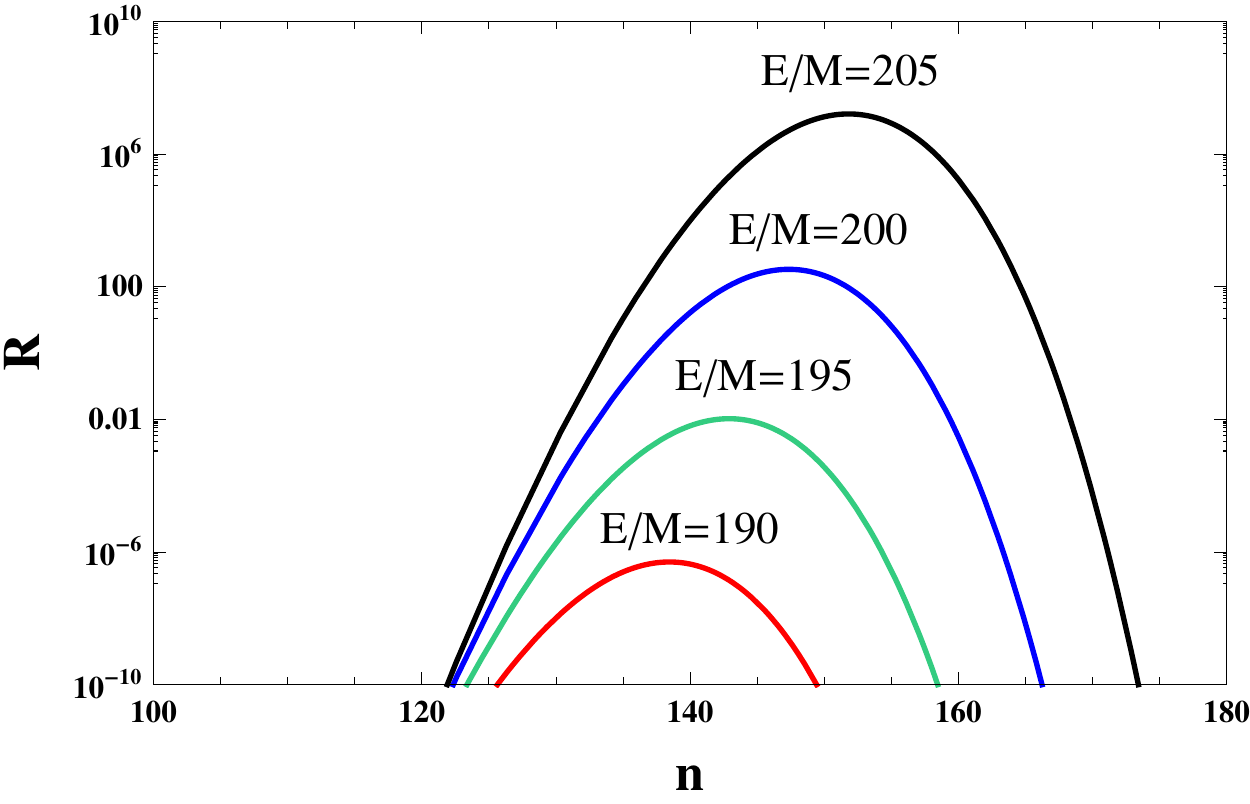}
\end{tabular}
\end{center}
\vskip-.6cm
\caption{Partial decay widths (in units of mass $M_h$) of a highly-energetic single-particle state into 
$n$ Higgs bosons $h$ plotted as function of $n$. The four lines correspond to the energies of the initial state equal 
190$M_h$, 195$M_h$, 200$M_h$ and 205$M_h$, as indicated. There is a sharp exponential dependence of the peak rate
on the energy varying from ${\cal R} \lesssim 10^{-6}$ at $E=190M_h$ (red line)  to ${\cal R} \gtrsim 10^{7}$ at $E=205M_h$ (black line). The peak multiplicities $n_\star \sim 150$ in these examples are not far from the maximally allowed values 
at the edge of the phase space $n_{\rm max} \sim E/M_h$.}
\label{fig:R}
\end{figure*}
 
 The resulting rates have a sharp exponential dependence on $n$ and, consequently, on energy.
 In order to be able to probe sufficiently high multiplicities, they have to be kinematically allowed, i.e. in our single-field example,
 $n < n_{\rm max} = E/M_h$. The amplitudes grow with $n$, as $\exp[n \log \lambda n]$,  reaching their maximal values in the soft limit where $n$ is
 maximal, but this effect is counter-acted by the diminishing phase-space volume near the edge of the kinematically accessible region.
 The competition between the two effects is clearly seen in the expressions for ${\cal R}$ already 
 at tree-level in Eq.~\eqref{eq:Rtreesimp}
and similarly in the re-summed perturbation theory expression in Eq.~\eqref{eq:Rnp}. The growth of the exponent
in ${\cal R}$ with increasing $\lambda n$ is counteracted at the edge of the phase-space by the
$\log \varepsilon$ factor where 
$\varepsilon = (E-M_h n)/(M_h n) \to 0$ when $n\to n_{\rm max}$. As a result we expect that the rate will peak at a 
non-perturbatively large value of $n \gg 1/\lambda$
but before the edge of the phase-space at $\varepsilon =0$ is reached.

 The relevant parameters are the energy
 $\sqrt{s} =E$ in the units of the elementary scalar mass, in our case $M_h$, and the number of particles in the final state
 $n$ rescaled by the (small) coupling constant, $\lambda n$. In the regime of relatively low-energies, $E/M_h \lesssim 10^2$,
 the multi-particle rates and cross-sections are exponentially small (essentially zero). But above the critical energy 
 $E_{\rm crit}$ in the region of $\simeq 200\, M_h$, using the plots in Fig.~\ref{fig:R} as a guide,  
 and for large values of $n$ towards the edge of the allowed phase-space, the exponential growth 
 in the rates starts competing with the 
 exponential suppression, the rates become of the order 1 and then blow up exponentially. 
 In Fig.~\ref{fig:R} we sketch the behaviour of the rate ${\cal R}$ in Eq.~\eqref{eq:Rnp} at fixed energies 
 $E=$190$M_h$, 195$M_h$, 200$M_h$ and 205$M_h$ as the function of the number of particles in the final state.
 For concreteness we have set $\lambda = 1/8$.
 The values of $E$ are chosen to illustrate the sharp rise in the rate from the exponentially suppressed to the
 exponentially enhanced level -- the transition which occurs very sharply with energy as it changes by just a few percent.  
 
 The structure of the peak in $n$ is easy to understand. Starting at $n_{\rm max}$ at the right of the plot, we are 
 at the end of the phase-space and
 the rate is zero. Then by decreasing the values of $n$ to the left of $n_{\rm max}$, the phase-space volume starts to grow
 and so does the rate ${\cal R}$. On the other hand, in the opposite limit, at low values of $n$, the rate is exponentially small again.
 Hence there must exist a local maximum, which clearly prefers as large as possible values of $n$ but before the edge of the
 phase-space is reached. 
 
 \bigskip
  
 In summary, we conclude that at sufficiently high energies $E>E_{\rm crit} \sim 2 \times 10^2 M_h$ (the precise value 
 would depend on the robustness of the model used\footnote{It is important to note that the overall structure of the 
 peaks observed in Fig.~\ref{fig:R} does not depend critically on the detailed form of the expression in \eqref{eq:Rnp}.
 All what is required is that the factorial growth of the tree-level amplitudes -- manifested as the $n \log \lambda n$ term 
 in the exponent of  \eqref{eq:Rnp} -- is not erased by the higher-order quantum corrections. The main points of the Higggsplosion and Higgspersion
 mechanisms discussed in this paper can be understood by simply assuming the behaviour of the type 
 sketched in Fig.~\ref{fig:R}.}) the multi-particle decay rates of an initial state into Higgs bosons
 develop a non-perturbative peak centred at $n=n_\star \gg 1/\lambda$ which tends to be near the edge of the kinematically
 accessible multi-particle phase-space, $n_\star = n_{\rm max} - \Delta n= E/M_h- \Delta n $. The peak 
 occurs in a non-perturbative regime, $n \gg 1/\lambda$, and the  width of the peak $2 \Delta n$
 is roughly of the order $1/\lambda$. Most of the energy available in the initial state is used to maximise the multiplicity 
 $n_\star$ of the final state bosons produced near the edge of the phase space, as such they correspond to relatively soft modes.
 
 It is tempting to interpret this peak as a creation of a semi-classical object -- a classicalon -- which then decays into 
 soft modes with ultra-high multiplicites. There are apparent parallels with the classicalization phenomenon
 \cite{Dvali:2010jz,Dvali:2010ns,Dvali:2016ovn} in which
 the theory prevents itself from probing shorter and shorter distances at very high energies by redistributing the 
 the energy of the initial state into many weakly interacting soft quanta.
 
\medskip
\section{Higgspersion, cross-sections and perturbative unitarity}
\label{sec:4}
\medskip

\begin{figure*}[h!]
\begin{center}
\includegraphics[width=0.6\textwidth]{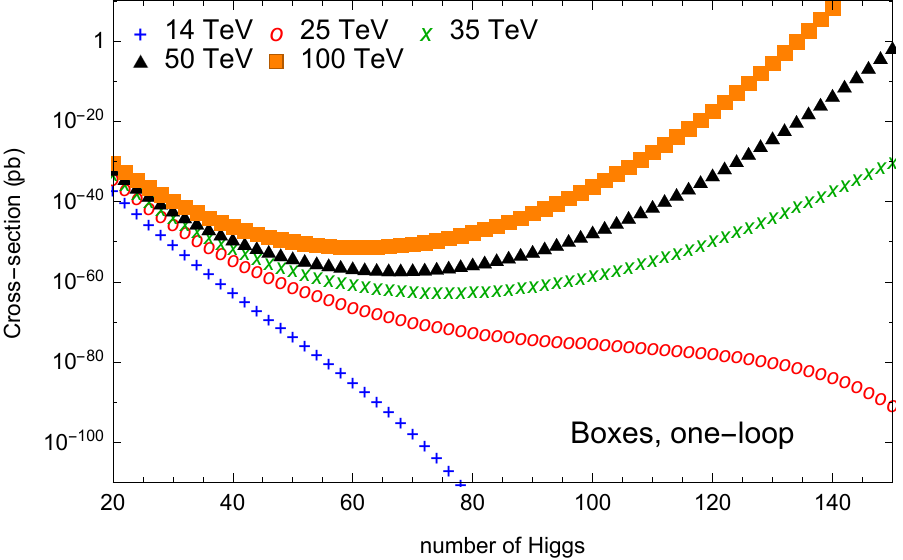}
\end{center}
\caption{Cross-sections for multi-Higgs production  at proton colliders including the PDFs for 
different energies of the proton-proton collisions plotted as the function of the Higgs multiplicity. Only the contributions from the boxes
are included. The Figure is taken from \cite{Degrande:2016oan}.}
\label{fig:pdf}
\end{figure*}

The scattering cross-sections for producing multiple Higgs bosons in the high-multiplicity limit $n\gtrsim 100$
at collider energies in the regime of 100 TeV were addressed and computed recently in Refs.~\cite{VVK3,Degrande:2016oan}
(with certain simplifying assumptions). These calculations consider the gluon fusion process
where intermediate highly energetic Higgs bosons are produced before subsequently branching
into high-multiplicity multi-Higgs final states. The results of Ref.~\cite{Degrande:2016oan} 
are based on the computation of the leading polygons -- the triangles, boxes, pentagons and hexagons --
to the gluon fusion production processes, further combined with the subsequent branchings to reach high final state multiplicities.
This can be represented as
\[
{\cal M}_{gg \to n\times h}\,=\, \sum_{\rm polygons} {\cal M}^{\rm polygons}_{gg \to k\times h^*}\,\,
 \sum_{n_1+\ldots+n_k=n}\, \prod_{i=1}^k \,{\cal A}_{h_i^* \to n_i\times h}\,,
\label{eq:chain}
\]
where the final partonic amplitude ${\cal M}_{gg \to k\times h}$ is convoluted with the gluon PDFs to obtain the
collider cross-section. The factors of  ${\cal A}_{h_i^* \to n_i\times h}$, 
after being squared and integrated over the multi-particle phase-space, result
in a factor of ${\cal R}_{n}(s)$ appearing in the cross-section. 
It was found that the characteristic energy and multiplicity scales where these cross-sections
become observable are within the  50 and 100 TeV regime with of order of 130 Higgses, or more, in the final state.
We refer the reader to Fig.~\ref{fig:pdf} 
(taken from Ref.~\cite{Degrande:2016oan}, and to the above reference for more details).

The Figure~\ref{fig:pdf} indicates that at collider energies below approximately 50 TeV, the processes are completely unobservable.
However at higher energies, from 50 to a 100 TeV, the cross-sections reach a picobarn level and become observable 
for 130 to 150 Higgs bosons produced. This is the regime where a dramatic change away from the usual weakly-coupled 
perturbative description of the electro-weak physics takes place. We note that the multiplicity range where the slope of 
the cross-section in  Fig.~\ref{fig:pdf}  changes so that the cross-section starts to increase with the multiplicity, corresponds to
the left-hand-side of the peaks in ${\cal R}$ shown in Fig.~\ref{fig:R}. The plot range in  Fig.~\ref{fig:pdf}  is cut-off before
the cross-sections for 35, 50 and 100 TeV reach their local maxima.

\bigskip
\subsection{Unitarity}
\bigskip

We will now argue that as soon as the cross-sections have reached the observable level, any
subsequent increase in the available energy will not result in the unbounded growth of the rates.
Instead, the cross-sections will actually decrease, and there will be no violation of perturbative unitarity.
For concreteness, consider the simplest process with a single intermediate off-shell Higgs 
propagator.\footnote{This corresponds to the contribution of triangle diagrams to the gluon fusion production.
The processes from all higher-order polygons, with more than one intermediate Higgs propagator can be
dealt with in a similar fashion.} The amplitude for this process reads ({\it cf.} Eq.~\eqref{eq:chain}):
\[
{\cal M}_{gg\to h^*} \,\times\, \frac{i }{p^2\,-\,M_h^2+i\, M_h\, \Gamma(p^2)} 
\,\times\, {\cal M}_{h^* \to n\times h} \,,
\label{eq:amp2}
\]
where  $\Gamma(s)$ is the energy-dependent total width of the Higgs at
the scale $s$, and it will lead to the Higgspersion of the total cross-section at asymptotically high energies.
In other words, the off-shell current ${\cal A}_{h^* \to n\times h}$
in Eq.~\eqref{eq:chain} includes the full dressed propagator times the amplitude ${\cal M}_{h^* \to n\times h}$.

In the limit $s \gg M_h^2, \, m_t^2$, the corresponding parton-level cross-section becomes,
\[ \sigma^{\Delta}_{gg \to n\times h} \, \sim \, 
y_t^{2}   
m_t^2  \log^4\left(\frac{m_t}{\sqrt{s}}\right) \,\times\, \frac{1}{s^2+M_h^4{\cal R}^2}
\,\times\, (2 \lambda)^{n-1}\, {\cal R}_n\,,
\label{eq:higgsper}
\]
and asymptotes to $1/{\cal R}$ in the limit ${\cal R}\to \infty$.
The inclusion of the decay width is of course only relevant when $\Gamma(s)$ becomes comparable to $s/M_h$. This conclusion is general and applies to higher-order polygons with more than one internal Higgs propagator.

\bigskip
In summary the multi-particle high-energy cross-section has the behaviour of the type,
\[
\sigma_{gg \to n\times h}\, \sim \, 
\begin{cases}
{\cal R} & :\,\,{\rm for}\,\,  {\cal R} \lesssim 1\\
1/{\cal R}\to 0 & :\,\,{\rm for}\,\,  {\cal R} \gg 1 \,\,{\rm at}\,\,s\to \infty\, .
\end{cases}
\label{eq:poly_ev_odd}
\]
The first line in the equation above is the result of Higgsplosion
and the second line is the consequence of the Higgspersion mechanism. 

\bigskip
\subsection{A comment on the K\"all\'en-Lehmann formula}
\bigskip

It can also be helpful to address  potential unitarity violations in the theory \cite{LRT,Jaeckel:2014lya}
using the K\"all\'en-Lehmann representation of the propagator for a scalar field $\phi$,
\[
\Delta_\phi (p)\,=\,\int_0^\infty \frac{ds}{2\pi} \, \frac{i}{p^2-m^2}\,\, \, \rho(s)\,,
\]
where $\rho(s)$ is the spectral density function, see e.g. \cite{Peskin},
\begin{eqnarray}
\nonumber
\rho(s)&=&\sum_{n}2\pi\,\delta\left(\sqrt{s}-\sum_{i=1}^{n} p_{i}\right)\,  \left |\left\langle 0|\phi(0) | n\right \rangle \right|^2
\,=\, 
2\pi\,Z_\phi\,\delta(s-m^2_{\phi})\,+\, \sum_{n\geq 2}\int d\Pi_n | {\cal A}(1^*\to n)|^{2}(s)
\\ \label{spectral}
&=&
2\pi\,Z_\phi\,\delta(s-m^2_{\phi})\,+\, \frac{1}{(s-m^2)^2}\,\sum_{n\geq 2}\int d\Pi_n | {\cal M}(1\to n)|^{2}(s)
\,,
\end{eqnarray}
and in the last line we have pulled out the external line propagators to represent the expression in terms
of the conventionally normalised scattering amplitudes ${\cal M}(1\to n)$. With this we find,
\[
\label{spec-rep} \Delta_\phi (p)\,=\,
\frac{i\,Z_\phi}{p^2-m^2}\,+\, i\,\sum_{n\geq 2} \int^{\infty}_{(nm)^2} \frac{ds}{2\pi}\,\, \frac{1}{p^2-s}\,\,\frac{\int d\Pi_n 
|{\cal M}(1\to n)|^{2}(s)}{(s-m^2)^2}\,.
\]
For $|p^2|<4m^{2}$ the second term on the right hand side gives a non-singular contribution to the propagator and the residue of the propagator pole is entirely determined by the first term.
The probability rates for
$1\to n$ processes  thus appear in Eq.~\eqref{spec-rep} as the additive order corrections to the propagator, and,
importantly, they are integrated over $s$. Thus it appears form this formula that if the multi-particle decay rates $\Gamma_s$
are exponentially divergent at large $s \gtrsim s_{\rm crit}$,
upon integration over $s$,  these corrections will blow up even at low values of $p^2$ , i.e. 
$|p^2|<4m^{2} \ll \sqrt{s_{\rm crit}}$.
Thus, 
it is tempting to say that to guarantee unitarity, the higher order terms in $n$ on the right hand side of Eq.~\eqref{spec-rep} should not be too large~\cite{LRT}.

\bigskip

This conclusion, however, depends on the validity of the above expression in \eqref{spec-rep}. 
Let us examine it and start by re-writing the term on the right hand side in terms of the imaginary part 
of the self-energy ({\it cf.}~\eqref{eq:2.10}-\eqref{eq:2.11}),
\[ - Z_\phi\,  {\rm Im}\, \Sigma(p^2) \,=\,\frac{1}{2} \int d\Pi_n |{\cal M}(1\to n)|^2 \,,
\]
so that,
\[\label{eq:4.9}
(-i)\Delta_\phi (p)\,=\,
\frac{Z_\phi}{p^2-m^2}\,+\,\sum_{n\geq 2} \int^{\infty}_{(nm)^2} \frac{ds}{2\pi}\,\, \frac{1}{s-p^2}\,\,
\frac{2\,Z_\phi\,  {\rm Im}\, \Sigma(s)}{(s-m^2)^2}\,.
\]
Note that the imaginary part of the self-energy is proportional to the discontinuity of the 
self-energy on the cut along the real axis of the complex variable $s$,
\[ {\rm disc}\, \Sigma(s)|_{s\ge (nm)^2)} \,=\, \Sigma(s+i \epsilon)-\Sigma(s-i\epsilon) \,=\, 2i\, {\rm Im}\,
\Sigma(s)\,, \]
hence the second term in \eqref{eq:4.9} is,
\[ \frac{1}{2\pi i} \sum_{n\geq 2} \int^{\infty}_{(nm)^2} ds\, 
 \frac{1}{s-p^2}\,\,
\frac{Z_\phi\,  {\rm disc}\,  \Sigma(s)}{(s-m^2)^2} \,.\nonumber\]
By adding to it an integral over the circular contour at $|s|\to \infty$ with the counter-clockwise orientation, we have ,
\begin{eqnarray}
&&\frac{1}{2\pi i} \sum_{n\geq 2} \int^{\infty}_{(nm)^2} ds\, 
 \frac{1}{s-p^2}\,\,
\frac{Z_\phi\,  {\rm disc}\,  \Sigma(s)}{(s-m^2)^2}\,+\,
\frac{1}{2\pi i}\oint_{|s|\to\infty} ds\, 
 \frac{1}{s-p^2}\,\,
\frac{Z_\phi\, \Sigma(s)}{(s-m^2)^2} \label{eq:disp}
\\
\qquad &&=\,
\frac{1}{2\pi i} \,\oint_{s=p^2} ds\, 
 \frac{1}{s-p^2}\,\,
\frac{Z_\phi\,\Sigma(s)}{(s-m^2)^2}\,\,=\,\, \frac{Z_\phi\,\Sigma(p^2)}{(p^2-m^2)^2} \nonumber
\,.
\end{eqnarray}
where on the last line we have used the Cauchy's theorem. Hence, we conclude that if the contour integral 
at $|s|\to \infty$ is negligible and can be added (that is if the integrand goes to zero at infinite $s$), the 
K\"all\'en-Lehmann formula for the renormalised propagator takes a familiar form:
\[\label{eq:KLconseq}
\Delta_\phi (p)\,=\,
\frac{i}{p^2-m^2}\,+\, \frac{i}{p^2-m^2}\, (-i\, \Sigma_R(p^2))\, \frac{i}{p^2-m^2}
\,,
\]
which is essentially a single perturbation in terms of the self-energy.

\bigskip

This derivation, however, breaks down completely when the ${\rm Im}\, \Sigma(s)$ explodes rather than falls off at $s\to \infty$,
which is precisely the case of interest for our consideration.
In this case the contour in \eqref{eq:disp} cannot be closed up at infinity and the dispersion relation \eqref{eq:disp} 
is invalid. We thus conclude that the formal justification of the perturbative K\"all\'en-Lehmann representation 
for the propagator in \eqref{spec-rep} or equivalently \eqref{eq:4.9} is meaningful only for a sufficiently 
well-behaved imaginary part of the self-energy expression at large $s$. When, on the other hand,
decay rates do not tend to vanish at infinity, one cannot use the dispersion relation to restore the real part from  
the imaginary part of the self-energy by closing up the contour, and the K\"all\'en-Lehmann representation 
in the form \eqref{spec-rep}, \eqref{eq:4.9} simply becomes invalid. Hence the growing multi-particle decay rates do not
necessarily imply the breakdown of unitarity of the theory. In the previous sub-section we have already argued 
that the relevant physical cross-sections in this case do not blow up and hence do not destroy the unitarity either.

\medskip
\section{Higgsplosion of heavy states below their mass-threshold}
\label{sec:5}
\medskip
 
 To outline the Higgsplosion approach as a solution to the Hierarchy problem in the Standard Model, let us consider a
 contribution of a hypothetical heavy scalar $X$ of mass $M_X$ to the Higgs boson mass parameter. We focus on the Lagrangian,
 \[
 {\cal L}_X \,=\, 
 \frac{1}{2} \partial^\mu X \,\partial_\mu X\,-\, \frac{1}{2} m_X^2 \, X^2 \,-\, 
 \frac{\lambda_{P}}{4}\, X^2 h^2\,.
\label{eq:toym2}
 \]
where $h$ is the Higgs boson.
 
 Calculating the contribution to the Higgs boson mass from the scalar $X$, we find
 \begin{eqnarray}
 \Delta M_h^2 &\sim& \lambda_P \int \frac{d^4p}{16\pi^4} \, \frac{1}{M_X^2\,-\, p^2\, +i\, {\rm Im}\, \Sigma_X(p^2)}
 \nonumber \\ \nonumber \\
 &=&  \lambda_P \int \frac{d^4p}{16\pi^4} \, \left( 
 \frac{M_X^2\,-\, p^2}{(M_X^2\,-\, p^2)^2\, +\, ({\rm Im}\, \Sigma_X(p^2))^2} 
 \,-\,  \frac{i\,{\rm Im}\, \Sigma_X(p^2)}{(M_X^2\,-\, p^2)^2\, +\, ({\rm Im}\, \Sigma_X(p^2))^2} 
 \right)
\nonumber 
\end{eqnarray}
Now, due to the Higgsplosion effect the multi-particle contributions to the width of $X$ 
explode at the values of the loop momenta $p^2 = s_\star$, where $\sqrt{s_\star} \simeq \mathcal{O}(25){\rm TeV}$
according to Fig.~\ref{fig:R}. This is much below
the masses of the hierarchically heavy states which we can assume to be at the GUT scale $\pm$ 10 orders of magnitude.
Because of the sharp exponential growth of the width ${\rm Im}\, \Sigma_X(s) \propto {\cal R}_n(s)$ with the energy,
it provides a sharp UV cut-off in the integral over the loop momenta at $p^2 =s_\star$.
Hence the integral in the expression above amounts to
\[ \Delta M_h^2 \,\propto\, \lambda_P \,\, \frac{s_\star}{M_X^2}\,\, s_\star\,.
 \label{eq:Xhp}
 \]
This is suppressed by the factor of 
$\left(\frac{\sqrt{s_\star}}{M_X}\right)^4 \,\simeq\, \left(\frac{25\,{\rm TeV}}{M_X}\right)^4$ relative to the 
normal expectations without the Higgsplosion-driven disintegration of the heavy particles.

\[ {\rm For}\quad \Gamma(s_\star) \simeq M_X \quad {\rm at} \quad s_\star \ll M_X^2 \quad \Longrightarrow \quad
\Delta M_h^2 \,\propto\, \lambda_P \,\, \frac{s_\star}{M_X^2}\,\, s_\star \,\ll\,  \lambda_P \, M_X^2\,.
\]

\bigskip

The reasoning above equally applies to any heavy modes, as far as they have a
non-vanishing interaction with the Higgs. These modes could be the heavy $10^{12}$ GeV sterile neutrinos 
which are important for the standard thermal Leptogenesis \cite{Fukugita:1986hr,Davidson:2002qv,Davidson:2008bu}, a heavy inflaton \cite{Guth:1980zm,Nanopoulos:1983up}, GUT-scale particles \cite{Georgi:1974sy,Fritzsch:1974nn}, flavons \cite{Froggatt:1978nt,Tsumura:2009yf},
or the 
heavy degrees of freedom that would appear at the $f_a \simeq 10^{11}$ GeV scale relevant for the axion \cite{Peccei:1977hh,Peccei:1977ur,Wilczek:1977pj,Weinberg:1977ma}.

\bigskip

At one-loop level, one can always estimate the contributions to the Higgs mass from the heavy states of any spin with 
generic interactions with the Higgs, using the Coleman-Weinberg effective potential,
\[ M_h^2 \,=\, \frac{\partial^2 V_{\rm eff}}{\partial h^2}\,,
\]
where
\[
V_{\rm eff} \,=\, \frac{1}{64\pi^2} \,\int^{\sqrt{s_\star}} d^4p\, \, {\rm STr} \, \log\left(p^2 +M_X(h)^2\right)\,,
\]
where ${\rm STr}={\rm Tr}(-1)^F$ is the supertrace, and $M_X(h)$ denotes the Higgs-field-dependent contribution to the 
heavy field mass in the $h(x)$ background. The main point, as above, is that the integral over the loop
momenta is cut-off at the relatively low scale $\sqrt{s_\star}$ where the Higgsplosion of the heavy states 
takes place.

It is remarkable that the Hierarchy problem introduced into the Standard Model by the existence of 
a microscopic light Higgs boson is addressed in this approach by Higgsploding the heavy states into the
original light Higgses. The underlying cause of the apparent problem provides its own solution.

\medskip
\section{Conclusions}
\label{sec:concl}
\medskip
The discovery of the Higgs boson, roughly 50 years after its prediction, marked one of the greatest successes of the SM. While its interactions with all other particles ensures the restoration of perturbative unitarity in  $2 \to 2$ scattering processes, it was long argued that the presence of a scalar particle in the theory could lead to unitarity violation in multi-Higgs production processes already at energies of $\mathcal{O}(100)$ TeV. Further, the Higgs boson, as an elementary scalar particle, suffers from the well-known Hierarchy problem. We have reexamined and connected both issues, thereby providing a simultaneous solution to both questions: We introduced the Higgsplosion mechanism, arguing that the rapid increase of the decay rate of very heavy or highly energetic particles is a physical effect, but that this effect leads to Higgspersion, i.e. it restores perturbative unitarity in multi-Higgs boson production processes. While the cross section of mutli-Higgs production processes can still reach observable levels, its exponential growth is avoided and the SM retains self-consistency to highest energies. 
Quantum corrections of heavy particles to the Higgs boson's mass are driving the Hierarchy problem. If however, the heavy particle's width increases rapidly beyond a certain energy threshold, these contributions are tamed and the Hierarchy problem can be avoided.

In summary, the existence of a microscopic light Higgs boson introduces arguably two fundamental issues to the Standard Model. However, we find that self-interactions of the Higgs boson provide mechanisms to heal the Standard Model and retain self-consistency of the theory. In case these mechanisms are realised in nature, it would be interesting to study their implications on explanations of fundamental questions in nature, e.g. the nature of dark matter or the underlying mechanism of the matter-antimatter asymmetry,

\bigskip

\section*{Acknowledgements}

We are grateful to Prateek Agrawal, Joerg Jaeckel, Michihisa Takeuchi and Tsutomu Yanagida for comments and useful discussions.
VVK thanks C{\'e}line Degrande, Joerg Jaeckel and Olivier Mattelaer for an earlier collaboration on related subjects and
acknowledges a Royal Society Wolfson Research Merit Award.
This work is supported by the STFC through the IPPP grant.%

\bigskip

\bibliographystyle{h-physrev5}

\end{document}